\documentclass{myaa}
\usepackage{graphicx}
\usepackage{txfonts}
\usepackage{natbib}
\usepackage{balance}
\usepackage[a4paper,breaklinks]{hyperref}
\idline{1}{1}
\begin{document}
\def\teff{$T\rm_{eff}$ }
\def\kms {\,$\mathrm{km\, s^{-1}}$ }
\def\kmss {\,$\mathrm{km\, s^{-1}}$}
\def\ms {$\mathrm{m\, s^{-1}}$ }

\newcommand{\Teff}{\ensuremath{T_\mathrm{eff}}}
\newcommand{\g}{\ensuremath{g}}
\newcommand{\gf}{\ensuremath{gf}}
\newcommand{\loggf}{\ensuremath{\log\gf}}
\newcommand{\glog}{\ensuremath{\log\g}}
\newcommand{\pun}[1]{\,#1}
\newcommand{\cobold}{\ensuremath{\mathrm{CO}^5\mathrm{BOLD}}}
\newcommand{\linfor}{Linfor3D}
\newcommand{\punms}{\mbox{\rm\,m\,s$^{-1}$}}
\newcommand{\punkms}{\mbox{\rm\,km\,s$^{-1}$}}
\newcommand{\abuhe}{\mbox{Y}}
\newcommand{\grav}{\ensuremath{g}}
\newcommand{\mlp}{\ensuremath{\alpha_{\mathrm{MLT}}}}
\newcommand{\mlpcm}{\ensuremath{\alpha_{\mathrm{CMT}}}}
\newcommand{\moh}{\ensuremath{[\mathrm{M/H}]}}
\newcommand{\senv}{\ensuremath{\mathrm{s}_{\mathrm{env}}}}
\newcommand{\shelio}{\ensuremath{\mathrm{s}_{\mathrm{helio}}}}
\newcommand{\smin}{\ensuremath{\mathrm{s}_{\mathrm{min}}}}
\newcommand{\spun}{\ensuremath{\mathrm{s}_0}}
\newcommand{\sstar}{\ensuremath{\mathrm{s}^\ast}}
\newcommand{\tauross}{\ensuremath{\tau_{\mathrm{ross}}}}
\newcommand{\ttaurelation}{\mbox{T$(\tau$)-relation}}
\newcommand{\Ysurf}{\ensuremath{\mathrm{Y}_{\mathrm{surf}}}}

\newcommand{\draftflag}{false}

\newcommand{\beq}{\begin{equation}}
\newcommand{\eeq}{\end{equation}}
\newcommand{\pdx}[2]{\frac{\partial #1}{\partial #2}}
\newcommand{\pdf}[2]{\frac{\partial}{\partial #2}\left( #1 \right)}

\newcommand{\var}[1]{{\ensuremath{\sigma^2_{#1}}}}
\newcommand{\sig}[1]{{\ensuremath{\sigma_{#1}}}}
\newcommand{\cov}[2]{{\ensuremath{\mathrm{C}\left[#1,#2\right]}}}
\newcommand{\xtmean}[1]{\ensuremath{\left\langle #1\right\rangle}}

\newcommand{\eref}[1]{\mbox{(\ref{#1})}}

\newcommand{\Vact}{\ensuremath{\nabla}}
\newcommand{\Vad}{\ensuremath{\nabla_{\mathrm{ad}}}}
\newcommand{\Veddy}{\ensuremath{\nabla_{\mathrm{e}}}}
\newcommand{\Vrad}{\ensuremath{\nabla_{\mathrm{rad}}}}
\newcommand{\Vraddiff}{\ensuremath{\nabla_{\mathrm{rad,diff}}}}
\newcommand{\cp}{\ensuremath{c_{\mathrm{p}}}}
\newcommand{\taueddy}{\ensuremath{\tau_{\mathrm{e}}}}
\newcommand{\vconv}{\ensuremath{v_{\mathrm{c}}}}
\newcommand{\Fconv}{\ensuremath{F_{\mathrm{c}}}}
\newcommand{\lmix}{\ensuremath{\Lambda}}
\newcommand{\Hp}{\ensuremath{H_{\mathrm{P}}}}
\newcommand{\Hptop}{\ensuremath{H_{\mathrm{P,top}}}}
\newcommand{\COBOLD}{{\sc CO$^5$BOLD}}

\newcommand{\changed}{}

\newcommand{\I}{\ensuremath{I}}
\newcommand{\Irot}{\ensuremath{\tilde{I}}}
\newcommand{\F}{\ensuremath{F}}
\newcommand{\Frot}{\ensuremath{\tilde{F}}}
\newcommand{\vsini}{\ensuremath{V\sin(i)}}
\newcommand{\vvsini}{\ensuremath{V^2\sin^2(i)}}
\newcommand{\vsinimu}{\ensuremath{\tilde{v}}}
\newcommand{\rotint}{\ensuremath{\int^{+\vsinimu}_{-\vsinimu}\!\!d\xi\,}}
\newcommand{\imu}{\ensuremath{m}}
\newcommand{\imupone}{\ensuremath{{m+1}}}
\newcommand{\nmu}{\ensuremath{N_\mu}}
\newcommand{\msum}[1]{\ensuremath{\sum_{#1=1}^{\nmu}}}
\newcommand{\wmu}{\ensuremath{w_\imu}}

\newcommand{\tchar}{\ensuremath{t_\mathrm{c}}}
\newcommand{\Nt}{\ensuremath{N_\mathrm{t}}}

\title{The forbidden 1082\pun{nm} line of sulphur:
\subtitle{the photospheric abundance
  of sulphur in the Sun and 3D effects}}

\author{
E. Caffau     \inst{1};
H.-G. Ludwig  \inst{2,1}
}
\institute{Observatoire de Paris-Meudon, GEPI, 92195 Meudon Cedex, France, 
\and
CIFIST Marie Curie Excellence Team
}
\authorrunning{Caffau et al.}
\titlerunning{[SI] line at 1082\pun{nm}}
\offprints{Elisabetta.Caffau@obspm.fr}
\date{Received ...; Accepted ...}

\abstract
{Sulphur is an element
which is formed in the $\alpha$-process and is easily
measured in the gaseous phase in external galaxies. Since
it does  not form dust, it is the preferred indicator
for $\alpha$-elements, rather than Si or Mg, for which
dust corrections are necessary.
The measurement of the sulphur abundance in stars is not an easy task,
relying mainly on high excitation lines with 
non-negligible deviations from LTE.
The 1082\pun{nm} sulphur forbidden transition is less sensitive to departures from
LTE and is less dependent on temperature uncertainties
than other sulphur lines usually employed as abundance indicators.
Therefore it should provide
a more robust abundance diagnostics.}
{To derive the solar photospheric abundance of sulphur from the 1082\pun{nm}
  [SI] line and to investigate 3D effects present in G- and F-type
  atmospheres at solar and lower metallicity.}
{High-resolution, high signal-to-noise solar intensity and flux spectra were
  used to measure the sulphur abundance from the [SI] 1082\pun{nm} line.
  \cobold\ hydrodynamical model atmospheres were applied to predict 3D
  abundance corrections for the [SI] line.}
{The solar sulphur abundance is derived to be 
  $7.15\pm (0.01)_{\rm stat}\pm (0.05)_{\rm sys}$, where the statistical uncertainty
  represents the scatter in the determination using four different solar
  spectra and the systematic uncertainty is due to the modelling of the blending lines.
  Sulphur abundances obtained from this line are insensitive to the
  micro-turbulence.  3D abundance corrections,
  found from strictly differential comparisons between 1D and 3D models,
  are negligible in the Sun, but
  become sizable for more metal-poor dwarfs.}
{}
\maketitle


\section{Introduction}

Weak forbidden lines arising from the ground level of an ion in the dominant
ionisation stage can be extremely useful for measuring abundances. They
are often unaffected by departures from LTE, and they are insensitive to the
atmospheric temperature {uncertainties} as well as micro-turbulence.  
On the other hand, forbidden lines are sensitive to gravity,
but this is of no concern in our analysis, since the solar gravity is well known.
The most widely used of
such forbidden lines is certainly the 630\pun{nm} [OI] line, which has often been
the oxygen abundance indicator of choice, even though it is blended with
a \ion{Ni}{i} line, which, in the solar spectrum, contributes about 1/3 of the
equivalent width (EW) of the feature.  Recently, \citet{ryde06} has used the
corresponding forbidden line of sulphur at 1082\pun{nm}, to measure the
sulphur abundance in a sample of stars in the Galactic disc. However, he did
not investigate the line in the solar spectrum.  The [SI] 1082\pun{nm} line is
clearly visible in solar spectra; we therefore decided to measure the solar sulphur
abundance from this line and to compare this measurement to values obtained
from other sulphur lines. We further investigated 3D effects influencing the [SI]
line with the help of \cobold\ hydrodynamical model atmospheres for
the Sun as well as for a few metal-poor G- and F-type stars.

To our knowledge, the first published analysis of the  [SI] 1082\pun{nm} line  
in the Sun was carried out by \citet{swensson68a}, who
related the transition
$3p^4$ $^3P_2$---$3p^4$ $^1D_2$ of \ion{S}{i},
to a feature  with an observed  wavelength
of 1082.12\pun{nm} that appeared  
next to a \ion{Cr}{i} line at 1082.164\pun{nm}
and a telluric H$_2$O line at 1082.215\pun{nm}. 
This identification was later  confirmed by \citet{swensson68b}.
He  estimated  an
EW of 0.35\pun{pm} (in \citealt{swensson68a}) and of 
$0.40\pm 0.05$\pun{pm} (in \citealt{swensson68b}) 
in the  solar observed intensity spectrum of \citet{delb63}.

Independently, \citet{swings69} identified the 1082\pun{nm} [SI] line in 
improved tracings of the solar spectrum and they published a convincing
detection of the solar [SI] line and solar sulphur abundance based on this line.
They measured the EW in the two spectra and found 
$0.32\pm0.02$\pun{pm} (from the Jungfraujoch solar intensity spectrum)
and $0.36\pm 0.03$\pun{pm} (from the solar scan of Kitt Peak), from which
they derived a sulphur abundance of 7.24 and 7.30, respectively,
by adopting a
\loggf\ of --8.61.  Using one-dimensional model atmospheres they predicted an
EW of 0.30\pun{pm} for the [SI] line at solar disc centre, 
for an assumed solar
sulphur abundance of 7.21 \citep{lambert68}.  This prediction is in good agreement with their
two measurements.


\section{Models and atomic data}

\ion{S}{i} is the dominant ionisation stage in the atmospheres considered in
this paper.  The [SI] line arises from the ground level, so that the abundance
of sulphur is fairly directly measured, instead of measuring the abundance
from an excited level representing a small fraction of the total population.

For the [SI] line we adopted
a laboratory wavelength 1082.1176\pun{nm} (in air)
and a \loggf\ of --8.617 
(see \href{http://physics.nist.gov/PhysRefData/ADS}{http://physics.nist.gov/PhysRefData/ADS}).
Our analysis is based on 3D model atmospheres
computed with the \cobold\ code \citep{wedemeyer03}.
More details about the models used can be found
in Caffau et al. (2007, in preparation).  In addition to the \cobold\ hydrodynamical
simulations we used several 1D models.  The spectral synthesis codes employed
are \linfor\ (see
\href{http://www.aip.de/~mst/Linfor3D/linfor_3D_manual.pdf}{http://www.aip.de/$\sim$mst/Linfor3D/linfor\_3D\_manual.pdf})
for all models 
and SYNTHE \citep{kurucz05} for the Holweger-M\"uller model. 
The 1D models we used are briefly described hereafter:
 
\begin{enumerate}
\item An ``internal 1D'' model generated by \linfor; this is a 1D atmospheric
  structure computed by horizontal and temporal averaging of a 3D model
  structure over surfaces of equal (Rosseland) optical depth. These 1D models
  provide estimates of the influence of fluctuations around the mean
  stratification on the line formation process. Comparing 3D and internal 1D
  models of this kind is largely independent of arbitrary input parameters,
  the micro-turbulence being the only free parameter.
\item In correspondence to each 3D \cobold\ model we constructed hydrostatic
  1D model atmospheres computed with the LHD code. LHD is a Lagrangian 1D
  (assuming plane-parallel geometry) hydrodynamical model atmosphere code. It
  employs the same micro-physics (equation-of-state, opacities) as \cobold.
  The convective energy transport is described by mixing-length theory. The
  spatial discretisation of the radiative transfer equation is similar to the
  one in \cobold, albeit simplified for the 1D geometry. A flux-constant hydrostatic
  stratification including radiative and convective energy transport processes
  is obtained by following
  the actual thermal and dynamical evolution of the atmosphere until a
  stationary state is reached. LHD produces standard 1D model atmospheres
  which are \textit{differentially comparable} to corresponding \cobold\ models.
  Remaining choices entering an LHD model calculation are the value of the
  mixing-length parameter, which formulation of mixing-length theory to use,
  and in what way turbulent pressure is treated in the momentum equation.
  Note, that these degrees of freedom are also present in other 1D model
  atmosphere codes.
\item A 1D solar model computed by F.~Castelli with version 9 of the
  ATLAS code and the solar abundances of \citet{sunabboasp} available at
  \href{http://wwwuser.oats.inaf.it/castelli/sun/ap00t5777g44377k1asp.dat}{http://wwwuser.oats.inaf.it/castelli/sun/ ap00t5777g44377k1asp.dat}
\item
The Holweger-M\"uller solar model \citep{hhsunmod, hmsunmod}.  
\end{enumerate}

A 3D \cobold\ model constitutes a statistical realisation of the atmospheric
flow field over a certain period of time. For spectral synthesis purposes, the
state of the flow is sampled at equal intervals in time. We informally
refer to a flow sample as a ``snapshot''. Each snapshot represents the stellar
photosphere at a particular instant of time.  Besides the 3D model
construction as such, 3D spectral synthesis also uses up quite a lot of computer time.
The computing time increases linearly with the total number of
snapshots~\Nt. Hence, \Nt\ should be restricted to as small a number as
possible, but chosen carefully to obtain a statistically meaningful result. We
like to emphasise that the strength of a line is usually a less critical
feature than its detailed line shape or line shift for which the velocity
field plays an important role.

Table~\ref{model3d} summarises the atmospheric parameters and snapshot
selection we did for our 3D models. Where possible, we tried to cover several
convective turn-over timescales in the surface layers (in the Sun
approximately 500\pun{s}). The sampling interval was chosen
not to be an integer fraction of the period of
the dominant oscillation mode of the computational box,
which is closely related to the period at which most power is expected for
stellar five-minute like oscillations. This ensures that one does not always
sample the oscillations at the same phase, which would lead to a bias in the
statistical properties of the velocity field controlling line broadening and
line shifts. Moreover, we selected our snapshots to ensure that the statistics
of the sub-sample closely followed the statistics of the whole ensemble, e.g.,
in terms of the resulting effective temperature and fluctuations around it.

Table~\ref{model3d} also lists a characteristic timescale $\tchar=\Hp/c$ (\Hp:
pressure scale height at $\tau_\mathrm{ross}=1$, $c$: sound speed), half the
(isothermal) acoustic cut-off period. From hydrodynamical models spanning a
large range in effective temperature, gravity, and metallicity 
\citet{fshgl}
found that the evolutionary
timescale of surface granulation roughly scales with \tchar. Hence,
considering \tchar\ one can relate an intrinsic timescale of the atmosphere to
the total time (column ``time'' in Table~\ref{model3d}) over which the
evolution of a relaxed model was followed.

As is obvious from Table~\ref{model3d}, the model with parameters 5770K/4.44/-2.0
does not fulfil the statistical criteria for the snapshot selection we outlined
before. However, since the line strength does not vary strongly over the
evolution of a 3D model we nonetheless felt that we could include this single
snapshot model to obtain at least an estimate of the abundance corrections for
its atmospheric parameters.

\begin{table}
\begin{center}
\caption{The \cobold\ models considered in this research.
The first three columns list the atmospheric parameters of the models,
the fourth column the number of snapshots~\Nt\ considered,
the last-but-one column the time interval covered by the
selected snapshots, and the last column a characteristic timescale~\tchar\ of
the evolution of the granular flow.
\label{model3d}}
\begin{tabular}{rrrrrr}
\hline
\noalign{\smallskip}
\Teff & \glog & [Fe/H] &\Nt & time & \tchar \\
(K)  & (cm s$^{-2}$)  &    &    & (s)    & (s)\\
\noalign{\smallskip}
\hline
\noalign{\smallskip}
5770 & 4.44 &   0.0 & 25 & 6000  & 17.8 \\
6500 & 4.00 &   0.0 & 28 & 16800 & 49.6 \\
5770 & 4.44 & --2.0 &  1 & 80    & 15.2 \\
6250 & 4.50 & --2.0 & 10 & 9600  & 15.9 \\
5900 & 4.50 & --3.0 & 19 & 9500  & 15.6 \\
6500 & 4.50 & --3.0 & 12 & 2400  & 16.0 \\
\noalign{\smallskip}
\hline
\end{tabular}
\end{center}
\end{table}


\section{Data}

We used two high-resolution, high signal-to-noise ratio, spectra of the centre
disc solar intensity. These were that of \citet{neckelobs} (hereafter
referred to as the  ``Neckel intensity
spectrum'') and that of Delbouille, Roland, Brault, \&\ Testerman (1981)
(hereafter referred to as the ``Delbouille intensity spectrum'')
(\href{http://bass2000.obspm.fr/solar_spect.php}{http://bass2000.obspm.fr/solar\_spect.php});
we further used the solar flux spectrum of \citet{neckelobs} and the one from
\href{http://kurucz.harvard.edu/sun.html}{http://kurucz.harvard.edu/sun.html}
(referred to as the ``Kurucz flux spectrum'').


\section{Data analysis}

\begin{figure}
\resizebox{7.5cm}{!}{\includegraphics[clip=true,angle=0]{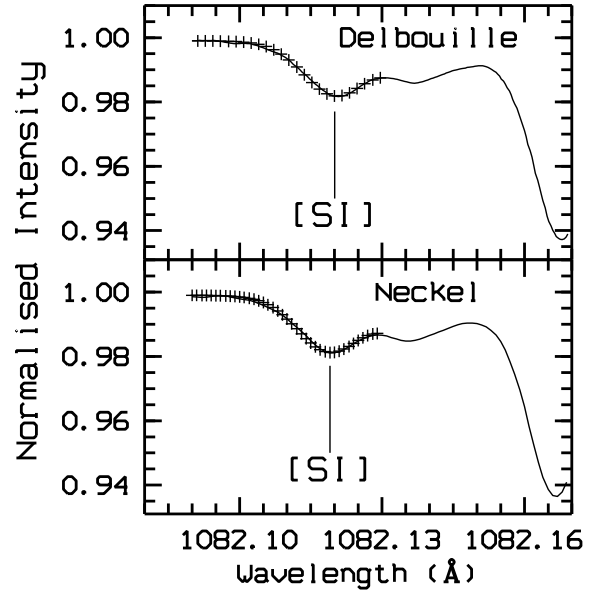}}
\caption{The observed intensity spectra (solid line) are plotted over
on the fit (crosses) obtained using a grid of synthetic
spectra based on the Holweger-M\"uller model computed
with SYNTHE.
Delbouille refers to the Delbouille intensity spectrum, Neckel to the Neckel
intensity spectrum (see text for details).}
\label{fitint}
\end{figure}

To derive  sulphur abundances in the solar spectra
we measured the equivalent width (EW)
and in addition fitted the line profiles.
EWs were measured using the IRAF task {\tt splot}
and the  deblending option, since the spectral range is rather complex.
The EW we obtain in this way takes into account the contribution
of the lines which lie to the
red (\ion{Cr}{i}, \ion{Fe}{i} and C$_2$) 
and to the blue  (\ion{Fe}{i} and C$_2$) of the [SI] line. 
For this reason our value is considerably lower than
the one found by \citet{swings69}.
However, since lines with poorly known atomic data crowd the region, 
we estimate an uncertainty in the
EW computation of at least 0.03\pun{pm}, which translates into an
error in the abundance of 0.05\,dex.
The line profile fitting was done using
a code, described in  \citet{zolfo05}, which performs
a $\chi ^2$ minimisation of the deviation between synthetic profiles
and the observed spectrum.
Figure~\ref{fitint} shows the fit of the solar intensities obtained
from a grid of synthetic spectra synthesised with SYNTHE and
based on the Holweger-M\"uller model.
 
\begin{table*}
\begin{center}
\caption{Solar sulphur abundances from the various observed spectra.
Col. (1) is the wavelength of the line followed
by an identification flag, D means Delbouille and N Neckel intensity spectrum,
F Neckel and K Kurucz flux spectrum.
Col.~(2) is the Equivalent Width.
Col.~(3)  is the sulphur abundance, A(S), according \cobold\ 3D model.
Cols.~(4)-(11) are the A(S) from 1D models, even numbered cols. correspond
to $\xi = 1.0$\kms, odd numbered cols. to $\xi = 1.0$\kms.
Col.~(12) is the A(S) from WIDTH code.
Col.~(13) is the A(S) from fitting with a HM grid;
col.~(14) and (15) are 3D corrections with
micro-turbulence of 1.5 and 1.0\kmss, respectively.
\label{sunzolfo}}
\begin{tabular}{rrrrrrrrrrrrrrr}
\hline
\noalign{\smallskip}
Wave      &  EW    & \multicolumn{10}{c}{A(S) from EW} & FIT  &\multicolumn{2}{c}{3D-1D} \\
nm        &  pm    & 3D   &  \multicolumn{2}{c}{1D} & \multicolumn{2}{c}{ATLAS} & \multicolumn{2}{c}{HM} & \multicolumn{2}{c}{LHD} &WIDTH & HM\\
          &        &      &   1.0  &   1.5  &   1.0  &   1.5  &   1.0  &   1.5  &   1.0  &   1.5  &       &         &  1.0    &   1.5 \\  
 (1)      &  (2)   & (3)  &  (4)   &  (5)   &  (6)   &  (7)   &  (8)   &  (9)   &  (10)  &  (11)  & (12)  & (13)    & (14)    & (15)\\
\noalign{\smallskip}
\hline
\noalign{\smallskip}
 1082D  & 0.212 & 7.140 & 7.140  & 7.139  & 7.154  & 7.153  & 7.170  & 7.170  & 7.123  & 7.122  &       &  7.150  &  0.0175 & 0.0182\\
 1082N  & 0.223 & 7.162 & 7.162  & 7.161  & 7.176  & 7.175  & 7.192  & 7.191  & 7.145  & 7.144  &       &  7.176  &  0.0175 & 0.0182\\
 1082F  & 0.258 & 7.138 & 7.148  & 7.147  & 7.173  & 7.172  & 7.180  & 7.179  & 7.137  & 7.135  & 7.133 &  7.168  &  0.0011 & 0.0024\\
 1082K  & 0.258 & 7.138 & 7.148  & 7.147  & 7.173  & 7.172  & 7.180  & 7.179  & 7.137  & 7.135  & 7.133 &  7.168  &  0.0011 & 0.0024\\
\noalign{\smallskip}
\hline
\end{tabular}
\end{center}
\end{table*}

The solar sulphur abundance derived from the [SI] line,
A(S)\footnote{A(S)=log (N(S)/N(H)) + 12}=$7.15\pm (0.01)_{\rm stat}\pm (0.05)_{\rm sys}$,
is in good agreement
with the one obtained from the 675\pun{nm} triplet in Caffau et al. (2007,
submitted to A\&A).  For the [SI] line the 3D abundance correction is small in the
Sun, and independent of the micro-turbulence.  The solar 
A(S) obtained
from this line (see Table~\ref{sunzolfo}), which is unaffected by 3D
corrections and is insensitive to NLTE effects, is believed to be close
to the actual solar abundance.

\begin{figure}
\resizebox{7.5cm}{!}{\includegraphics[clip=true,angle=0]{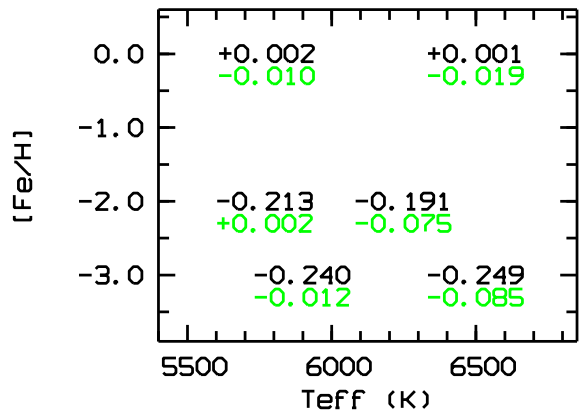}}
\caption{3D abundance corrections: each correction is located according the
parameters (Teff and [Fe/H]) of the related model; the models have very
similar gravity (\glog\ in between 4.44 and 4.50). 
Black numbers depict the 3D-1D~LHD differences, grey (green in colour)
ones the 3D-1D internal model differences.}
\label{cor3d}
\end{figure}

\begin{figure}
\resizebox{7.5cm}{!}{\includegraphics[clip=true,angle=0]{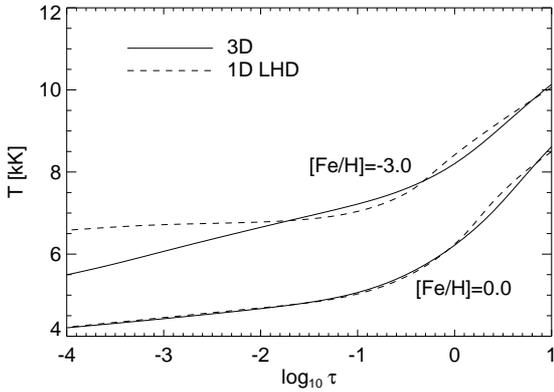}}
\caption{Average temperature structures of 3D \cobold\ (solid lines) in
  comparison to 1D LHD (dotted lines) models. The lower pair of curves is for
  5770K/4.44/0.0 (solar) models, the upper pair for 5900K/4.50/-3.0 models.
  For clarity, the later models were shifted by +2000\pun{K}.}
\label{ttau}
\end{figure}

As a theoretical exercise we computed 3D corrections for flux spectra in
metal-poor stars.  As expected, the [SI] line is not sensitive to the
micro-turbulence, the 3D corrections do not depend on this parameter.  The
results are depicted in Fig.~\ref{cor3d}.  From the plot we can deduce that for 
solar-metallicity stars the 3D abundance corrections are negligible. The corrections
increase for decreasing stellar metallicity. They become sizable
at [Fe/H]=--2.0 and [Fe/H]=--3.0.
If we compare the average temperature structure of the 3D models to the 1D
reference LHD models, we realise that they are very close for solar-metallicity 
stars (see Fig.~\ref{ttau}).  For metal-poor stellar models the 3D
and 1D~LHD average temperature structures are not in good agreement.  At
$\log\tau<-1$, 3D models are cooler, and the difference increases towards
external layers.  According to the contribution function the [SI] line is
formed mostly at $\log\tau\approx -1$ for solar-metallicity stars, but at
$\log\tau\approx -2$ and $\log\tau\approx -3$, for [Fe/H] of --2.0 and --3.0,
respectively.  
The basic reason for the behaviour of the contribution function is that 
the continuous opacity (mostly H$^-$) decreases for lower
metallicity at fixed Rosseland optical depth in the outer atmospheric
layers. The effect is more 
pronounced in 3D models due to the lowering of the average temperature
at low metallicity.
Since the line emerges from a transition from the ground level,
the population of the line's lower level is not very sensitive to the absolute
temperature, so that temperature differences between 3D and 1D models are not
so important.  However, the temperature \textit{gradients} also differ
markedly between 3D and 1D models in the line-forming layers. In the weak line
approximation, the line strength is proportional to the gradient of the (log)
source function on the continuum optical depth scale, so EWs computed with the
3D model with a steeper temperature gradient tend to be larger than the ones
from a 1D model. In comparison to 1D a lower sulphur abundance is needed in a
3D model to obtain the same equivalent width of the line, corresponding to a
negative 3D abundance correction.
 
For metal-poor F-type stars contributions due to horizontal fluctuations
(3D-1D internal) are about one third of the overall abundance corrections, while
for metal-poor G-type stars this contribution is negligible.  
Horizontal temperature fluctuations in metal-poor photospheres
are significantly smaller in G-type than in F-type atmospheres. The reason
does not lie so much in a different level of the convective dynamics between
the two spectral types. The reason is related to the temperature attained due
to the cooling by convective overshooting. In G-type stars photospheric
temperatures become so low that $\mathrm{H}_2$ molecular formation sets in. This
drives up the specific heat, leading to smaller changes of the temperature in
response to pressure disturbances induced by overshooting motions. As evident
from Fig.~\ref{cor3d}, the ``demarcation line'', where this effect becomes
important, is located around \mbox{\Teff$\approx$6000\pun{K}} for dwarf stars.

\section{Discussion}

Our analysis implies that the [SI] line constitutes a robust
indicator of the sulphur abundance, in spite of the fact that
the measurement of EW is affected by a rather large error.
Our A(S)=$7.15\pm (0.01)_{\rm stat}\pm (0.05)_{\rm sys}$
is consistent with the value provided by \citet{lodders}, A(S)=$7.21\pm 0.05$.

\balance
It is of some interest to consider the 3D corrections for this line.  A 3D
model atmosphere differs from a 1D one in essentially two respects. In the
first place, a hydrodynamical simulation provides a mean temperature
structure which is different from that of a 1D atmosphere assuming radiative
equilibrium; this is illustrated in Fig.~\ref{ttau}.  In the second place,
the hydrodynamical simulation takes into account the effects of horizontal
temperature fluctuations, which is something a 1D model cannot do. Bearing
this in mind it is fairly easy to understand the meaning of the two values
for the correction provided in Fig.~\ref{cor3d}.  The 3D-1D corrections
(grey/green in the plot) is the difference between the 3D model and the
internal 1D model, which is the average of the 3D model over surfaces of
equal (Rosseland) optical depth.  Therefore, this correction provides
information about the influence of the horizontal fluctuations on the derived
abundance.  As suggested from the figure, this correction increases as the
effective temperature increases, but it is always very small when compared
to the global correction.  The 3D-1D LHD model correction (black in the
plot) takes into account both the different mean structure of the two
models, as well as the horizontal fluctuations.  The LHD model, that employs
the same micro-physics (equation-of-state, opacities) as \cobold, provides a
better estimate of the ``3D effects'' than the use of other 1D model
atmospheres (e.g. ATLAS models). The effects of different assumptions about
the micro-physics may result in a cancellation or enhancement of the effects,
which are related to the different nature of the 3D model.

For the Sun the 3D abundance corrections are negligible. For metal-poor stars
the full 3D corrections become sizable,
the sulphur abundance is lowered by a factor of 1.6, 
and they are dominated by effects related
to the mean temperature profile, rather than by that of the horizontal
fluctuations.  Due to the lack of available models we cannot give 3D
corrections for slightly metal-poor stars yet, but we intend to do so as soon
as we have these models at hand.


\begin{acknowledgements}
The authors wish to thank Rosanna Faraggiana and Piercarlo Bonifacio for
  their help and suggestions with this research. The authors acknowledge gratefully financial
  support from EU contract MEXT-CT-2004-014265 (CIFIST).
\end{acknowledgements}

\bibliographystyle{aa}

\end{document}